\documentclass[12pt]{article}

\usepackage{graphicx}

\begin{document}

\begin{center}
{\bf Inflation of universe due to nonlinear electrodynamics} \\
\vspace{5mm} S. I. Kruglov
\footnote{E-mail: serguei.krouglov@utoronto.ca}

\vspace{3mm}
\textit{Department of Chemical and Physical Sciences, University of Toronto,\\
3359 Mississauga Road North, Mississauga, Ontario L5L 1C6, Canada} \\
\vspace{5mm}
\end{center}

\begin{abstract}
A model of nonlinear electrodynamics with a dimensional parameter $\beta$ is considered. Electromagnetic fields are the source of the gravitation field and inflation of the universe. We imply that the universe is filled by stochastic magnetic fields. It is demonstrated that after the universe inflation the universe decelerates approaching the Minkowski spacetime. We evaluate the spectral index, the tensor-to-scalar ratio, and the running of the spectral index which approximately agree with the PLANK and WMAP data.
\end{abstract}

\vspace{3mm}


\vspace{3mm}


\vspace{3mm}

\section{Introduction}

One of the way to explain the universe inflation is to modify general relativity by introducing $F(R)$ gravity. The choice of $F(R)$ functions is not unique, and therefore, there are many modified gravity models \cite{Capozziello}, \cite{Nojiri}. At the same time, in early time of the universe evolution,
the electromagnetic fields were very strong and quantum corrections should be taken into account \cite{Jackson}. Therefore, it is natural to modify classical electrodynamics for strong electromagnetic fields. Born and Infeld \cite{Born} proposed nonlinear electrodynamics (NLED) that smoothing singularity of point-like charges and gives the finite value of self-energy.
Classical electrodynamics with loop (quantum) corrections becomes NLED \cite{Heisenberg}, \cite{Schwinger}, \cite{Adler}.
NLED coupled to the gravitation field can describe inflation \cite{Garcia}, \cite{Camara}, \cite{Elizalde}, \cite{Novello3}, \cite{Novello}, \cite{Novello1}, \cite{Kruglov7}, \cite{Kruglov8}, \cite{Kruglov9}.
In this work we explore a new NLED model which is an effective model of electromagnetic fields and is valid for strong fields, and at weak fields it is converted into Maxwell's electrodynamics. Thus, the correspondence principle holds. We will demonstrate that the universe with the stochastic magnetic background, within our NLED, undergos the universe inflation.

The structure of the paper is as follows. In section 2 we describe a new model of NLED with a dimensional parameter $\beta$. The cosmology of the universe with stochastic magnetic field background is studied in Sec. 3. We show that there are not singularities of the energy density, pressure and the Ricci scalar. In Sec. 4 the universe evolution is investigated and we find the scale factor as a function of time. We study also the causality and a classical stability of the model. In Sec. 5 we evaluate the spectral index $n_s$, the tensor-to-scalar ratio $r$, and the running of the spectral index $\alpha_s$ that are in approximate agreement with the PLANK and WMAP data. Sec. 6 is devoted to a conclusion.

The units with $c=\varepsilon_0=\mu_0=1$ and the metric signature $\eta=\mbox{diag}(-,+,+,+)$ are used.

\section{The model of NLED}

Consider NLED with the Lagrangian density
\begin{equation}
{\cal L} = -\frac{{\cal F}}{(\beta{\cal F}+1)^2},
\label{1}
\end{equation}
where $\beta$ is a dimensional parameter, ${\cal F}=(1/4)F_{\mu\nu}F^{\mu\nu}=(\textbf{B}^2-\textbf{E}^2)/2$, $F_{\mu\nu}=\partial_\mu A_\nu-\partial_\nu A_\mu$ is the field strength tensor. The symmetric energy-momentum tensor is given by
\begin{equation}
T_{\mu\nu}=\frac{(\beta{\cal F}-1)F_\mu^{~\alpha}F_{\nu\alpha}}{(1+\beta{\cal F})^3}
-g_{\mu\nu}{\cal L},
\label{2}
\end{equation}
with the non-vanishing trace
\begin{equation}
{\cal T}\equiv T_\mu^{~\mu}=\frac{8\beta{\cal F}^2}{(1+\beta{\cal F})^3}.
\label{3}
\end{equation}
Putting $\beta\rightarrow 0$ we arrive at classical electrodynamics, ${\cal L}\rightarrow -{\cal F}$ and trace (3) vanishes, ${\cal T}\rightarrow 0$. Because the energy-momentum tensor trace is not zero the scale invariance is broken due to the dimensional parameter $\beta$.
 The energy density $\rho$ and the pressure $p$ obtained from Eq. (1) are given by
\begin{equation}
\rho=-{\cal L}-E^2\frac{\partial {\cal L}}{\partial{\cal F}}=\frac{(1-\beta{\cal F})E^2}{(1+\beta{\cal F})^3} +\frac{{\cal F}}{(1+\beta{\cal F})^2},
\label{4}
\end{equation}
\begin{equation}
p={\cal L}+\frac{E^2-2B^2}{3}\frac{\partial {\cal L}}{\partial{\cal F}}=-\frac{{\cal F}}{(\beta{\cal F}+1)^2}+\frac{(E^2-2B^2)(\beta{\cal F}-1)}{3(\beta{\cal F}+1)^3}.
\label{5}
\end{equation}

\section{Cosmology}

The action of general relativity (GR) coupled with NLED, described by Eq, (1), reads
\begin{equation}
S=\int d^4x\sqrt{-g}\left[\frac{1}{2\kappa^2}R+ {\cal L}\right],
\label{6}
\end{equation}
where $\kappa^{-1}=M_{Pl}$, $M_{Pl}$ is the reduced Planck mass, and $R$ is the Ricci scalar. Electromagnetic fields are the source of gravitational fields. The Einstein and electromagnetic field equations, found by varying action (6), are
\begin{equation}
R_{\mu\nu}-\frac{1}{2}g_{\mu\nu}R=-\kappa^2T_{\mu\nu},
\label{7}
\end{equation}
\begin{equation}
\partial_\mu\left(\frac{\sqrt{-g}F^{\mu\nu}(\beta{\cal F}-1)}{(\beta{\cal F}+1)^3}\right)=0.
\label{8}
\end{equation}
Consider homogeneous and isotropic cosmological spacetime with the line element
\begin{equation}
ds^2=-dt^2+a(t)^2\left(dx^2+dy^2+dz^2\right),
\label{9}
\end{equation}
with $a(t)$ being a scale factor.
We suppose that stochastic magnetic fields are the cosmic background with the wavelength smaller than  the curvature.
Averaging the magnetic fields, which are sources in GR \cite{Tolman}, give the isotropy of the Friedman-Robertson-Walker (FRW) spacetime.
The averaged magnetic fields obey equations as follows:
\begin{equation}
<\textbf{B}>=0,~~~~<E_iB_j>=0,~~~~<B_iB_j>=\frac{1}{3}B^2g_{ij}.
\label{10}
\end{equation}
The brackets $<>$ denote an average over a volume larger than the radiation
wavelength and smaller as compared with the curvature of spacetime. For simplicity we omit below the brackets $<>$ . The energy-momentum tensor of NLED obeying Eqs. (10) is represented by a perfect fluid \cite{Novello1}.
The Friedmann equation is given by
\begin{equation}
3\frac{\ddot{a}}{a}=-\frac{\kappa^2}{2}\left(\rho+3p\right).
\label{11}
\end{equation}
The dots over the letter denote the derivatives with respect to the cosmic time.
The acceleration of the universe holds when $\rho + 3p < 0$.
According to the standard cosmological models a symmetry in the direction holds, and as a result $<B_i> = 0$.
Making use of Eqs. (4) and (5) we obtain (for $\textbf{E}=0$)
\begin{equation}
\rho+3p=\frac{4B^2(2-3\beta B^2)}{(2+\beta B^2)^3}.
\label{12}
\end{equation}
The plot of the function $\beta(\rho+3p)$ versus $\beta B^2$ is represented in Fig. 1.
\begin{figure}[h]
\includegraphics[height=4.0in,width=4.0in]{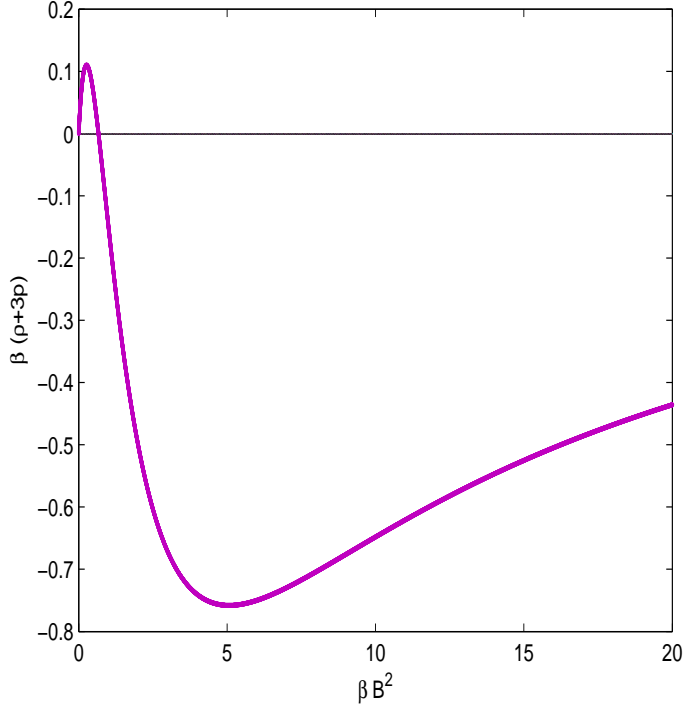}
\caption{\label{fig.1}The function  $\beta(\rho+3p)$ vs. $\beta B^2$. }
\end{figure}
The acceleration of the universe occurs ($\rho + 3p < 0$) when $B>\sqrt{2}/\sqrt{3\beta}\simeq 0.82/\sqrt{\beta}$. The strong magnetic fields result the universe inflation.
The conservation of the energy-momentum tensor, $\nabla^\mu T_{\mu\nu}=0$, gives the relation
\begin{equation}
\dot{\rho}+3\frac{\dot{a}}{a}\left(\rho+p\right)=0.
\label{13}
\end{equation}
From Eqs. (4) and (5), for the case $\textbf{E} = 0$, we find
\begin{equation}
\rho=\frac{2 B^2}{\left(2+\beta B^2\right)^2},~~~~\rho+p=\frac{8B^2(2-\beta B^2)}{3\left(2+\beta B^2\right)^3}.
\label{14}
\end{equation}
By virtue of Eqs. (14), after integration of Eq. (13), one obtains
\begin{equation}
B(t)=\frac{B_0}{a^2(t)}.
\label{15}
\end{equation}
Here $B_0$ is the magnetic field corresponding to the scale factor $a(t)=1$.
Thus, the scale factor increases due to inflation and the magnetic field decreases. From Eqs. (14) and (15) we find the energy density and pressure depending on the scale factor
\begin{equation}
\rho(t)=\frac{2a^4(t) B_0^2}{\left(2a^4(t)+\beta B_0^2\right)^2},
~~~~p(t)=\frac{2a^4(t)B_0^2(2a^4(t)-7\beta B_0^2)}{3\left(2a^4(t)+\beta B_0^2\right)^3}.
\label{16}
\end{equation}
From Eqs. (16) we find
\begin{equation}
\lim_{a(t)\rightarrow 0}\rho(t)=\lim_{a(t)\rightarrow 0}p(t)=\lim_{a(t)\rightarrow \infty}\rho(t)=\lim_{a(t)\rightarrow \infty}p(t)=0.
\label{17}
\end{equation}
Thus, singularities of the energy density and pressure at $a(t)\rightarrow 0$ and $a(t)\rightarrow \infty$ are absent. The plot of the equation of state (EoS) $w=p(t)/\rho(t)$ versus $x=[2/(\beta B_0^2)]^{1/4}a(t)$ is represented in Fig. 2.
\begin{figure}[h]
\includegraphics[height=4.0in,width=4.0in]{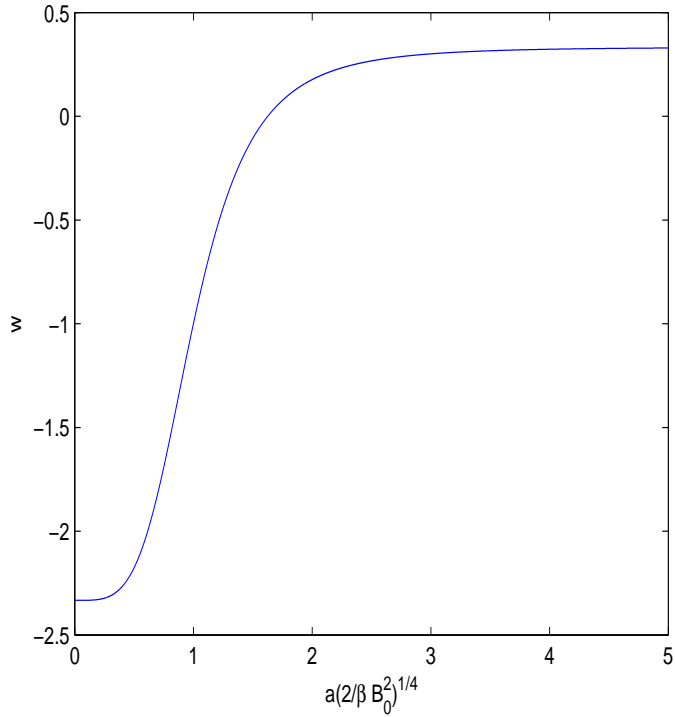}
\caption{\label{fig.2}The function  $w$ vs. $[2/(\beta B_0^2)]^{1/4}a$.}
\end{figure}
Making use of Eqs. (16) one obtains
\begin{equation}
\lim_{x\rightarrow\infty} w=\frac{1}{3}.
\label{18}
\end{equation}
At $a(t)\rightarrow \infty$ we have the EoS for ultra-relativistic case \cite{Landau}.
At $x=1$, we have EoS corresponding to de Sitter spacetime $w=-1$.
One obtains the curvature from Einstein's equation (7) and from the trace of energy-momentum tensor (3),
\begin{equation}
R=\kappa^2{\cal T}=\frac{16\kappa^2\beta B^4}{(2+\beta B^2)^3}=\kappa^2\left[\rho(t)-3p(t)\right].
\label{19}
\end{equation}
The plot of the function $\beta R/\kappa^2$ versus $[2/(\beta B_0^2)]^{1/4}a$ is given in Fig. 3.
\begin{figure}[h]
\includegraphics[height=4.0in,width=4.0in]{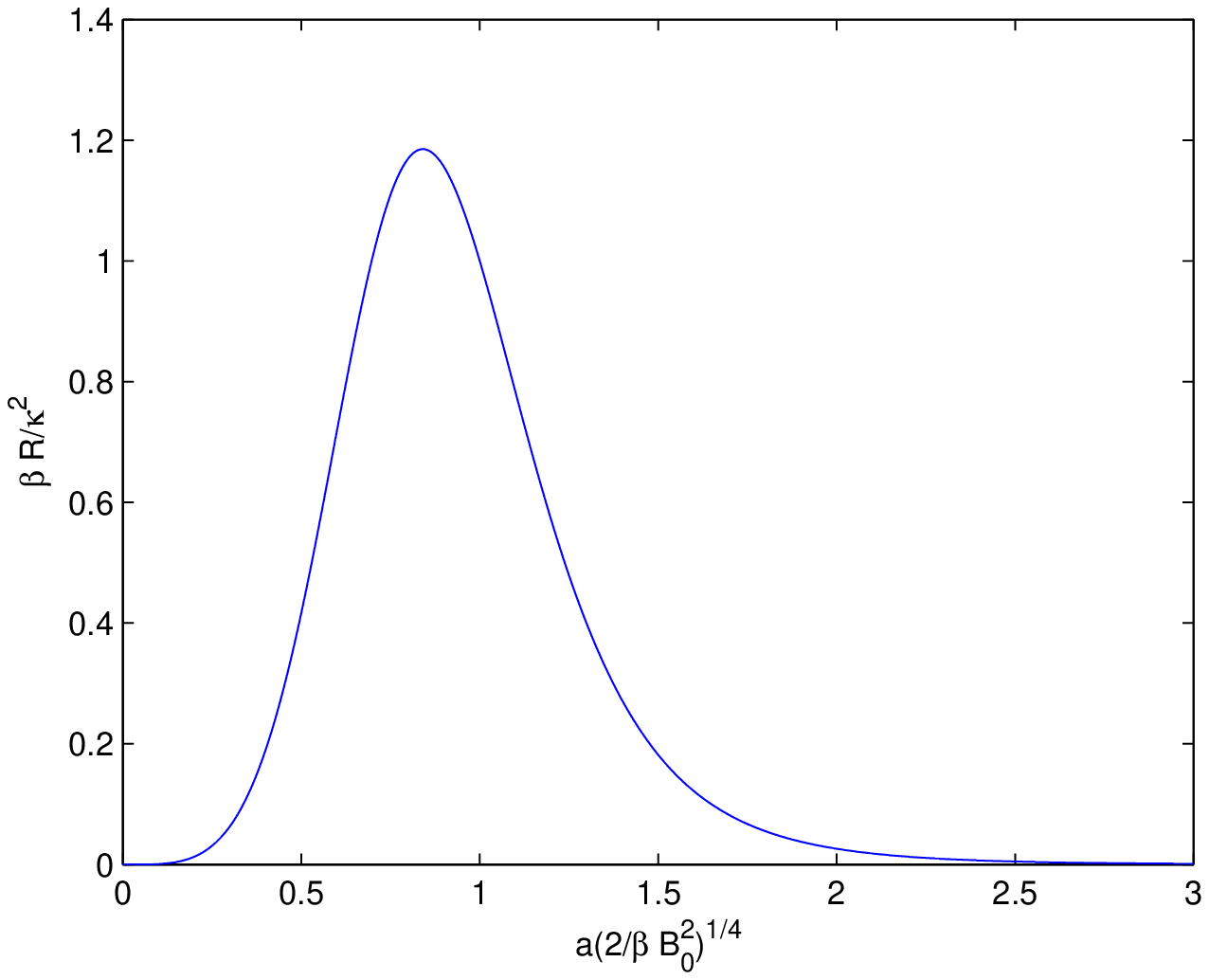}
\caption{\label{fig.3}The function  $\beta R/\kappa^2$ vs. $[2/(\beta B_0^2)]^{1/4}a$. }
\end{figure}
By virtue of Eqs. (17) and (19) we make a conclusion that
\begin{equation}
\lim_{a(t)\rightarrow 0}R(t)=\lim_{a(t)\rightarrow \infty}R(t)=0.
\label{20}
\end{equation}
Thus, the singularity of the Ricci scalar is absent. The Ricci tensor squared $R_{\mu\nu}R^{\mu\nu}$ and the Kretschmann scalar $R_{\mu\nu\alpha\beta}R^{\mu\nu\alpha\beta}$ can be expressed as linear combinations of $\kappa^4\rho^2$, $\kappa^4\rho p$, and $\kappa^4p^2$ \cite{Kruglov7} and they, according to Eq. (17), vanish at $a(t)\rightarrow 0$ and $a(t)\rightarrow \infty$.
At $t\rightarrow\infty$
the scale factor increases and spacetime becomes Minkowski's spacetime. It follows from Eqs. (12) and (15) that the universe accelerates at $a(t)<(3\beta)^{1/4}\sqrt{B_0}/2^{1/4}\simeq 1.1\beta^{1/4}\sqrt{B_0}$, i.e. the universe inflation occurs.

\section{ Evolution of the universe}

To find the dependance of the scale factor on time we explore the second Friedmann equation for three dimensional flat universe
\begin{equation}
\left(\frac{\dot{a}}{a}\right)^2=\frac{\kappa^2\rho}{3}.
\label{21}
\end{equation}
Making use of Eqs. (14) and (15), from Eq. (21) we obtain
\begin{equation}
\dot{a} =\frac{\kappa ba^3}{\sqrt{3\beta}(a^4+b^2)},~~~~b\equiv \frac{\sqrt{\beta}B_0}{\sqrt{2}}.
\label{22}
\end{equation}
After integrating Eq. (22) we come to the quadratic equation
\begin{equation}
a^4-\frac{2\kappa bt}{\sqrt{3\beta}}a^2-b^2=0.
\label{23}
\end{equation}
We took the integration constant such that $a(0)=\sqrt{b}$. The real solution to Eq. (23) is given by
\begin{equation}
a(t)=\sqrt{\frac{\kappa bt}{\sqrt{3\beta}}+\sqrt{\left(\frac{\kappa bt}{\sqrt{3\beta}}\right)^2+b^2}}.
\label{24}
\end{equation}
The scale factor increases in the time from acceleration phase to deceleration phase.
Introducing the variable $x=\kappa bt/\sqrt{3\beta}=\kappa B_0t/\sqrt{6}$, Eq. (24) becomes
\begin{equation}\label{25}
a(x)=\sqrt{x+\sqrt{x^2+b^2}}.
\end{equation}
It follows from Eq. (25) that at $x<b/\sqrt{3}$ ($t<\sqrt{\beta}/\kappa$) the universe accelerates and at $x>b/\sqrt{3}$ ($t>\sqrt{\beta}/\kappa$) the universe decelerates.
The solution (24) at $t\rightarrow\infty$ becomes $a\rightarrow 2^{1/4}\sqrt{\kappa B_0t}/3^{1/4}$ corresponding to the radiation era. The scale factor $a(t)$ approaches zero only at $t\rightarrow -\infty$.
Thus, the model describes inflation and there are not singularities at the early epoch.
The plot of the function  $a(x)$  is represented in Fig. 4.
\begin{figure}[h]
\includegraphics[height=4.0in,width=4.0in]{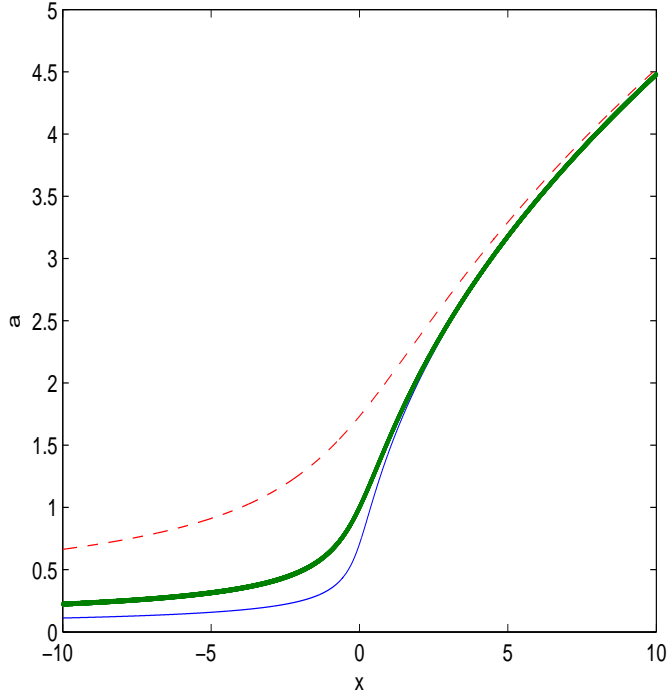}
\caption{\label{fig.4}The function  $a$ vs. $x=\kappa B_0t/\sqrt{6}$. Solid thin line corresponds to $b=0.5$,  solid thick line - to $b=1$, and dashed line - to $b=3$.}
\end{figure}

\subsection{ Sound Speed and Causality}

The causality occurs if the speed of the sound is less than the local light speed, $c_s\leq 1$ \cite{Quiros}. A classical stability requires that the square sound speed is positive, $c^2_s> 0$.  From Eqs. (4) and (5) we find speed squared at $E=0$
\begin{equation}
c^2_s=\frac{dp}{d\rho}=\frac{dp/d{\cal F}}{d\rho/d{\cal F}}=\frac{7\beta^2 B^4-32\beta B^2+4}{3(4-\beta^2 B^4)}.
\label{26}
\end{equation}
A classical stability holds at $c^2_s> 0$, which gives
\begin{equation}
\sqrt{\beta} B<\sqrt{\frac{2(8-\sqrt{57})}{7}}\simeq 0.36~~or~~\sqrt{2}<\sqrt{\beta} B<\sqrt{\frac{2(8+\sqrt{57})}{7}}\simeq 2.1.
\label{27}
\end{equation}
The causality, $c_s\leq 1$, occurs at
\begin{equation}
\sqrt{\beta} B\leq\sqrt{2}~~or~~\sqrt{\beta} B\geq\sqrt{\frac{2(4+\sqrt{21})}{5}}\simeq 1.85.
\label{28}
\end{equation}
The causality and classical stability hold at $\sqrt{\beta}B<\sqrt{2(8-\sqrt{57})/7}$ or $\sqrt{2(8+\sqrt{57})/7}>\sqrt{\beta} B\geq\sqrt{2(4+\sqrt{21})/5}$.
At $\sqrt{\beta}B>\sqrt{2}/\sqrt{3}\simeq 0.82$ the universe inflation occurs. Thus, the required bound $c_s^2\leq 1$ takes place at the deceleration phase, $\sqrt{\beta}B<\sqrt{2(8-\sqrt{57})/7}\simeq 0.36$. The plots of the function $c_s^2$ vs. $[2/(\beta B_0^2)]^{1/4}a$ are represented in Figs. 5 and 6.
\begin{figure}[h]
\includegraphics[height=4.0in,width=4.0in]{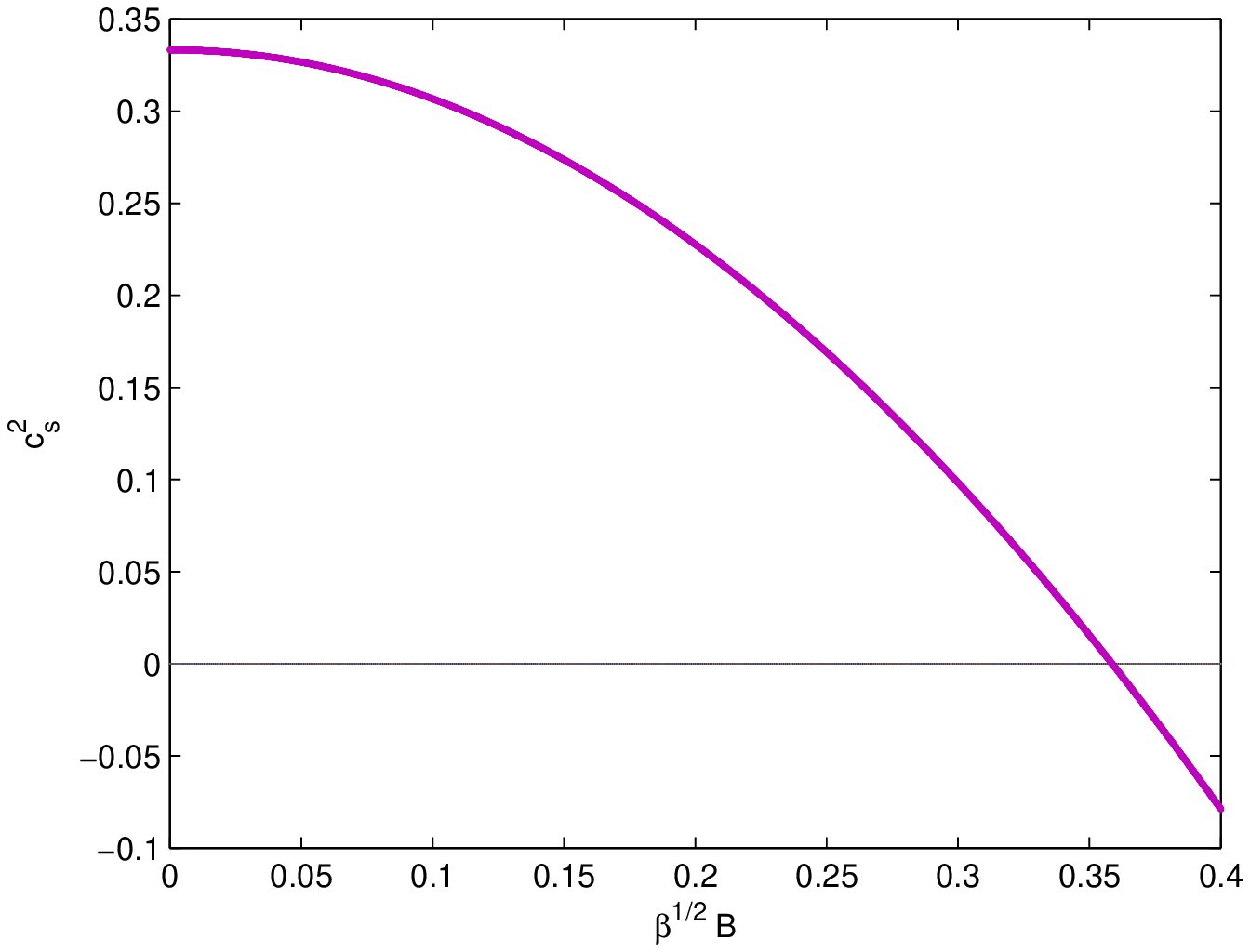}
\caption{\label{fig.5}The function $c_s^2$ vs. $\sqrt{\beta} B$.}
\end{figure}
\begin{figure}[h]
\includegraphics[height=4.0in,width=4.0in]{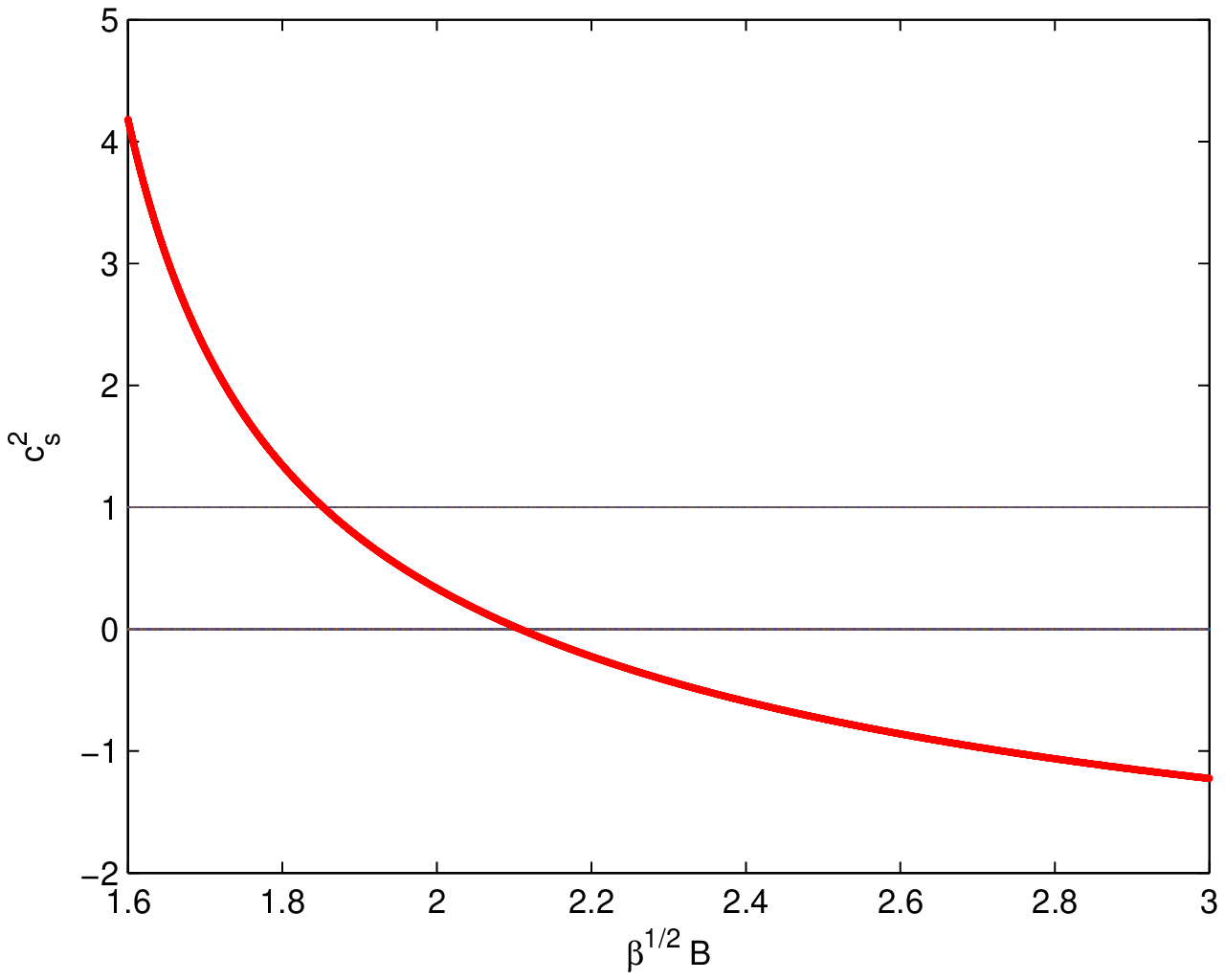}
\caption{\label{fig.6}The function $c_s^2$ vs. $\sqrt{\beta} B$.}
\end{figure}

\section{Cosmological parameters}

From Eqs. (4) and (5) we find (for $\textbf{E}=0$)
\begin{equation}
p=-\rho-\frac{4\rho(\beta B^2-2)}{3(\beta B^2+2)},
\label{29}
\end{equation}
\begin{equation}\label{30}
 \beta B^2=\frac{1-2\beta\rho-\sqrt{1-4\beta\rho}}{\beta\rho}.
\end{equation}
It follows from Eq. (4) that maximum of the energy density occurs, $\rho_{max}=1/(4\beta)$, at $B=\sqrt{2/\beta}$. From Eqs. (29) and (30) one obtains EoS for the perfect fluid
\begin{equation}
p=-\rho+f(\rho),~~~~f(\rho)=-\frac{4\rho\left(1-4\beta\rho-\sqrt{1-4\beta\rho}\right)}{3\left(
1-\sqrt{1-4\beta\rho}\right)}.
\label{31}
\end{equation}
The plot of the function $p\beta$ versus $\rho\beta$ is given in Fig. 7.
\begin{figure}[h]
\includegraphics[height=4.0in,width=4.0in]{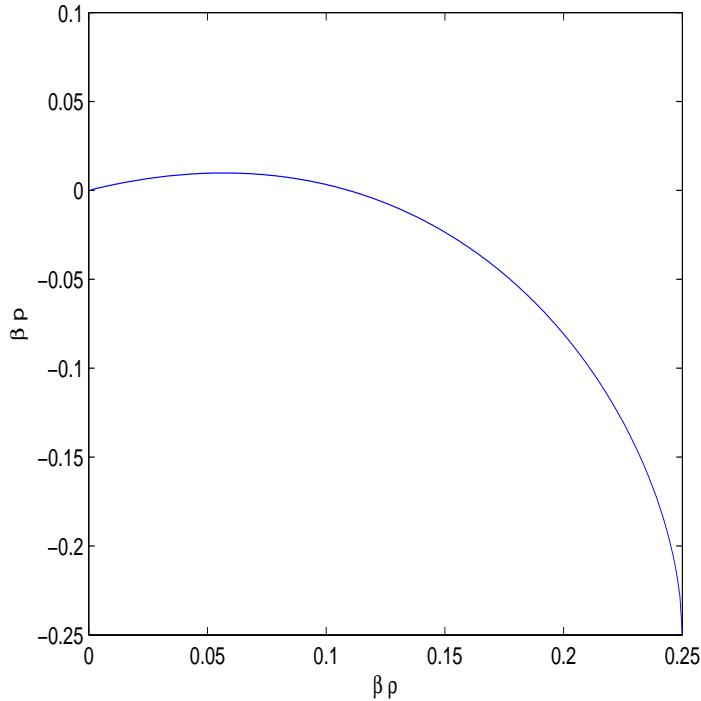}
\caption{\label{fig.7}The function $p\beta$ vs. $\rho\beta$.}
\end{figure}
When the condition $|f(\rho)/\rho|\ll 1$ holds during the inflation, the expressions for the spectral index $n_s$, the tensor-to-scalar ratio $r$, and the running of the spectral index $\alpha_s=dn_s/d\ln k$ read \cite{Odintsov}
\begin{equation}
n_s\approx 1-6\frac{f(\rho)}{\rho},~~~r\approx 24\frac{f(\rho)}{\rho},~~~\alpha_s\approx -9\left(\frac{f(\rho)}{\rho}\right)^2.
\label{32}
\end{equation}
 From Eq. (32) we obtain the relations
\begin{equation}
r=4(1-n_s)=8\sqrt{-\alpha_s}=\frac{32\left(\sqrt{1-4\beta\rho}+4\beta\rho-1\right)}{
1-\sqrt{1-4\beta\rho}}.
\label{33}
\end{equation}
The PLANCK experiment \cite{Ade} and WMAP data \cite{Komatsu}, \cite{Hinshaw} give the results
\[
n_s=0.9603\pm 0.0073 ~(68\% CL),~~~r<0.11 ~(95\%CL),
\]
\begin{equation}
\alpha_s=-0.0134\pm0.0090 ~(68\% CL).
\label{34}
\end{equation}
When we take the value $r=0.13$ one obtains from Eqs. (33) the values for the spectral index $n_s=0.9675$ and the running of the spectral index $\alpha_s=-2.64\times 10^{-4}$.
From Eq. (33) we obtain the value $\beta\rho\simeq 0.2499$ corresponding to these reasonable values of
cosmological parameters. It should be mentioned that the maximum of the energy density is $\rho_{max}=0.25/\beta$.
The energy density gives the value of the magnetic field $B\simeq \sqrt{2}/\sqrt{\beta}$ which corresponds to the inflation phase.

\section{Conclusion}

We have considered the magnetic universe with a stochastic background, $<B^2>\neq 0$, in the framework of new model of nonlinear electromagnetic fields.
The model for homogeneous and isotropic cosmology can describe the universe inflation. We show that singularities of the energy density, pressure, the Ricci scalar, the Ricci tensor squared, and the Kretschmann scalar are absent. A stochastic magnetic field is the source of the universe inflation at the early epoch, and then at $B> \sqrt{2}/\sqrt{3\beta}$, the universe decelerates approaching to the radiation era.
We obtained the range of magnetic fields when the classical stability and the causality hold.
The spectral index, the tensor-to-scalar ratio, and the running of the spectral index calculated are in approximate agreement with the PLANK and WMAP data.

It should be mentioned that in the early models with scalar fields there was a problem
with the graceful exit. This means that once inflation started it never ends, i.e. inflation is eternal. Then ”new” and ”chaotic” models of inflation appeared \cite{Linde} to provide a graceful exit from inflation. There are other ways to solve the graceful exit problem: to modify general
relativity by $F(R)$ gravity, to introduce fields that nonminimally couple to gravity
and there are other roads \cite{Nojiri}. In our model of inflation the problem of
a graceful exit is absent. According to Figs. 1 and 5 after inflation the universe decelerates approaching to the radiation era.
Only the problem of the current acceleration exists. To solve this problem one may
modify general relativity or to introduce fields coupled nonminimally with
gravity. The possible generalization of the theory under consideration is to introduce nonminimal
coupling gravity with NLED. This was done for Maxwell's electrodynamics coupled with $F(R)$ gravity
in \cite{Odintsov1}.

\end{document}